\documentclass[sigconf]{acmart}

\usepackage{booktabs} 
\usepackage{multirow}
\usepackage{algpseudocode}
\usepackage{calrsfs}
\usepackage{balance}
\usepackage{enumitem}
\usepackage{romannum}
\usepackage{graphicx}
\usepackage[linesnumbered]{algorithm2e}
\usepackage{pgfplots}

\renewcommand\footnotetextcopyrightpermission[1]{} 
\begin{document}

\fancyhead{}


\settopmatter{printacmref=false}

\title{Scalable Panel Fusion Using Distributed Min Cost Flow}

\newcommand{\superscript}[1]{\ensuremath{\sqrt{\sqrt{•}}^{\textrm{#1}}}}
\def\sharedaffiliation{\end{tabular}\newline\begin{tabular}{c}}
\def\wu{\superscript{*}}
\def\wp{\superscript{+}}
\def\wg{\superscript{\dag}}

\author{Swapnil Shinde}
\affiliation{
	\institution{Comscore}
}
\email{sshinde@comscore.com}

\author{Jukka Ranta}
\affiliation{
	\institution{Comscore}
}
\email{jranta@comscore.com}

\author{Paul Deitrick}
\affiliation{
	\institution{Comscore}
}
\email{pdeitrick@comscore.com}

\author{Matthew Malloy}
\affiliation{
	\institution{University of Wisconsin, Comscore}
}
\email{matthew.malloy@wisc.edu}

\renewcommand{\shortauthors}{ et al.}

\begin{abstract}
Modern audience measurement requires combining observations from disparate \emph{panel} datasets.  Connecting and relating such panel datasets is a process termed \emph{panel fusion}.  This paper formalizes the panel fusion problem and presents a novel approach to solve it.   We cast the panel fusion as a network flow problem, allowing the application of a rich body of research.  In the context of digital audience measurement, where panel sizes can grow into the tens of millions, we propose an efficient algorithm to partition the network into sub-problems.  While the algorithm solves a relaxed version of the original problem, we provide conditions under which it guarantees optimality. We demonstrate our approach by fusing two real-world panel datasets in a distributed computing environment. 

\end{abstract}

%
%



\maketitle

\section{Introduction} \label{sec:introduction} Audience measurement -- the estimation of the size and characteristics of an audience -- plays a fundamental role in the advertising ecosystem: advertisers pay content producers (for example -- newspapers, websites, and television networks) based on the number of people exposed to content and advertisements. 
 
Audience measurement has traditionally relied on audience \emph{panels}. An audience panel is a group of participants that agree to logging of their exposure to media, and demographic and household information. Some media -- such as over the air television -- require traditional panels for measurement, as there is no return communication path.  Measurement of media delivered to internet connected devices follows a distinct paradigm -- information can be collected electronically, through a variety of means -- monitoring software, web page, video, and mobile application tracking \emph{tags}, and internet connected televisions.  The result is extremely large scale panels with narrow purview.  Instead of measuring \emph{all} behavior of a person, only a  single medium is measured.  This is the challenge of \emph{cross-media} audience measurement. 
 
In practice, multiple large scale \emph{disparate} panels are required to provide accurate statistics on cross-media audiences.  Consider the following example.  Imagine a first panel measuring mobile application consumption, and a second panel measuring website consumption, with no way to directly link the panelists.  How can one measure the size of the audience \emph{shared} by an app and a website?  \emph{Panel fusion} addresses this question by relating panelists from two or more panels, and combining their observed behaviors to create accurate cross-media statistics. 

Panel datasets represent selected groups of people or households with demographic attributes, \emph{e.g.,} age,  gender,
household size, income, race, and ethnicity. More advanced demographic attributes like supermarket preferences, automobile
ownership, product purchase interests can also be associated with panelists.  Each panelist is assigned a \emph{projection weight}. These weights are used to make the overall panel represent the demographic composition and behavior of the audience \emph{universe}, using techniques such as Raking \cite{deville1993generalized}.  Different panels are comprised of different panelists, each with a different bias and different associated projection weight. A key component to a successful fusion is matching the panelists in the different panels in a way that maintains the composition and behavior present in both panels in the combined dataset in an optimal way.  The alignment of projection weights between two disparate panels can be cast as a network flow problem with the weights of the first panel representing the \emph{source}, and the weights of the second panel representing the \emph{sink}.  The cost of associating the behavior of a panelist from the first panel with the behavior of a panelist from the second panel is defined by a subjective measure of similarity between their behavior and demographic information.  

 
In this paper, we address the problem of combining two disjoint panel datasets by casting the problem as a \emph{min-cost-circulation} network problem.  The formulation guarantees projection weight alignment between the panels, and optimizes to prefer associations that align demographic profiles and observed behavior similarities.  The approach and solution can be applied to any panel fusion problems that need to be optimally solved at a large scale. We provide a computationally efficient algorithm designed for distributed compute platforms. 

We demonstrate our methodology at internet scale with large datasets from Comscore, an internet and TV audience measurement and analytics company. Comscore's digital network consists of web pages-, advertisement- and application tags deployed on websites and advertisements across the internet. The scale of the dataset is immense, consisting of more than 50 billion measured events each day.  These measurements are organized to define a single large scale (census) inferred panel of over 30 million panelists.  The breadth of the information in this large scale panel is limited in that only the activity on tagged websites, apps and advertisements is observed. Comscore also maintains complementary traditional panel datasets.  These panels are comprised of panelists who have agreed to install monitoring software on their desktop, mobile or other digital devices. The monitoring software reports on the online behavior of the panelist, creating a complete picture of their internet activity.  These panel datasets consists of approximately one million panelists with demographic and behavioral information.  After casting the panel fusion problem as a minimum cost circulation problem, we fuse the two datasets, creating a single source panel, with scale derived from the first panel, and breadth derived from the second. 

In summary, this paper makes the following contributions.  First, we formalize the panel fusion problem, and cast it as a minimum cost circulation problem, following the work of \cite{Soong2001THEAO}.  Next, we propose a scalable solution to solve the fusion problem, and more general minimum cost circulation problems.  We demonstrate our results at scale, and report on the characteristics of the fused panels.

\section{Problem definition} \label{sec:methods} In this section, we define the panel fusion problem.  We cast the problem as a transportation problem, which is a special case of the more general network flow problem.  

Let $U$ represent a set of panelists with $|U| = n_1$, and $V$ be a second set of panelists with $|V| = n_2$.  Each panelist $i$ is a node in a bipartite graph, $G = (U,V,E)$, with $i \in  \{ U \cup V \}$.  The two sets of panelists are disjoint, $\{ U \cap V \} = \{ \}$.  Each panelist $i$ has an associated projection weight, $w_i \in \mathbb{R}$, $w_i > 0$, and a feature vector, $z_i$ of length $m$.   
 Edges between nodes $i$ and $j$,  denoted  $e_{i,j} \in E$ have an associated \emph{cost} per unit flow $c_{i,j}$, which depends on the dissimilarities of the features of the panelist: $c_{i,j} = d(z_i, z_j )$, where $d(\cdot, \cdot)$ is a measure of distance between the attributes between the panelist.  

In our instantiation of the problem, the feature vector contains categorical values corresponding to demographic categories (i.e, Male, age 20-24, etc.), and real valued features representing minutes associated with types of behavior (i.e, one hour spent on social media). Additionally, the panelist weights, specified a priori, are used to scale the panelist behavior to match the measurement universe \footnote{For example, if the panel is comprised of a random sample from the general population at a rate of 1/1000, each panelist is given a weight of 1000.}.   Both panels represent the same measurement universe, hence 
\begin{equation} \label{eq:1}
\sum_{i \in U} w_i = \sum_{j \in V} w_j
\end{equation}
is given.     


To provide intuition, a match between panelist $i \in U$ and panelist $j \in V$ is considered appropriate if the following qualitative constraints are met: 
\begin{enumerate}[label=(\Roman*)]
\item the projection weights of matched panelists are the same, and
\item feature vectors $z_i$ and $z_j$ are as similar as possible. 
\end{enumerate}
We make these qualities precise in Section \ref{sec:transport}.  Assignment constraints as defined above imply a one to one assignment between panelists.  In practice, our approach requires fractional assignment of panelists, as in general, $n_1 \neq n_2$.  


\subsection{Panel fusion as transportation problem} \label{sec:transport}
                               
 The aim of the minimum cost flow problem is to determine a path with the least cost through a network, while simultaneously satisfying supply and demand constraints of the nodes.  In the panel fusion problem, a dense bipartite graph where the projection weights represent the \emph{source} -- the supply -- and \emph{sink} -- the demand -- of the network. Let edge $e_{i,j}$ be have upper capacity bound = $\infty$ and lower capacity bound = 0. The upper bound is considered $\infty$ for simplicity of explanation but can be changed to influence flow where needed. In many practical applications, the lower bound is set to 0 to allow discarding certain edges completely if needed. Let the flow on an $e_{i,j}$ be given by  $x_{i,j} $.  The minimum cost flow problem is written as follows: \par

\begin{equation}
\begin{aligned}
\min_{x} \quad & \sum_{i,j} c_{i,j}  x_{i,j} \\
\textrm{s.t.} \quad & \sum_{ j \in V} x_{i,j} = w_i  \mbox{ for all } i  \in U  \\
  &\sum_{i \in U} x_{i,j} = w_j  \mbox{ for all } j \in V   \\
\end{aligned}
\end{equation}
The minimum cost circulation problem aims to minimize the total cost by adjusting the flow between the nodes, $x_{i,j}$.  In other words, the sum of the costs (the product of the cost per unit flow, and the flow), across all edges in the bipartite graph, is minimized. 

The constraints state that, for every panelist $i \in U$, the fractional weights of the panelists in $V$ that flow to/from $i$ must sum to $w_i$.  The same constraint applies for every panelist $j$ in $V$.  The constraints guarantee alignment of the panelist weights as proposed in constraint \Romannum{1}.

\section{Algorithm} \label{sec:methods} 
 
We present the core framework for solving panel matching and then work through natural modifications which lead to the practical implementation of a solution with large scale distributed platforms. Section 5 explains more about minimum circulation flow problem and related algorithms.
     
\begin{algorithm}
\DontPrintSemicolon
\SetNoFillComment 
\KwData{\\
$U, V \ \mbox{-- disjoint sets of  panelist}$\\
$\qquad w_i,  i \in U $ and $w_j, j \in V$  -- projection weights\\
$\qquad z_i,  i \in U $ and $z_j, j \in V$  -- feature vectors\\
}

\KwResult{\\
$x_{i, j}$ for $i \in U$ and $j \in V$ -- assigned flow for edge $e_{i, j}$
}
\Begin{
Normalize \ features \ and \ calculate \ distance \\
$\qquad$ between $\ i \in U \ and \ j \in V$\\
Generate bipartite graph $G = (U, V, E)$\\
$\qquad$ with $e_{i,j} = d(z_i, z_j)$  for $(i, j) \in U\times V$\\
\tcc{cost scaling successive approximation}
\it{graphSolution = MCFSolver( $ G$)} \\
$\mathit{assignedPairs = generateAssignedPairs(graphSolution) }$

\KwRet{$assignedPairs$};
}
\caption{Core fusion algorithm framework}
\label{alg:algorithm1}
\end{algorithm}

Algorithm \ref{alg:algorithm1} outlines the core fusion algorithm with a single bipartite graph. Line [2-3] describes the normalization of features and calculation of distances between panelists of the two disjoint panel sets. Normalization and distance methods are not explained for the sake of simplicity but any effective method can be applied. \par
Line 4 describes generation of bipartite graph as defined in Sec 2.1. $\mathit{MCFSolver}$ on line 6 contains the \emph{cost scaling successive approximation algorithm} to solve with minimum cost flow in bipartite graph.
$\mathit{generateAssignedPairs}$ on line 7 transforms the flow solution into assigned pairs of ($i$, $j$, $x_{(i, j)}$) where $i$ is matched with $j$ with flow of $x_{(i, j)}$.  \par
As defined in \cite{ahuja1995applications}, the \emph{mass balance constraint} and the \emph{flow bound constraints} of network flow problem asserts our Constraint \Romannum{1}, which enforces that the weights of matched pair weights are perfectly aligned with outgoing/incoming flows.
\emph{Cost minimization} in network flow solution where cost is a function of the distance between panelists optimizes the results against Constraint \Romannum{2}.

\subsection{Node splitting}
It is worth mentioning that the solution includes fractional assignments. Therefore, a given node can be assigned to multiple nodes while making sure that split or fractional flow is bounded by supply or demand. We refer this case as $\mathit{node \ splitting}$. In the panel fusion problem, this can be interpreted as one panelist being  matched with multiple panelists by distributing its projection weight while following Equation 2.\par
This is a side effect of solving panel fusion with a single bipartite graph. We will provide an optimization where this can be completely avoided if needed. This optional constraint can be enforced depending on the requirements of the fusion problem at hand.


\subsection{Algorithm optimization}
For many applications in modern audience measurement, the above algorithm requires solving a large bipartite graph.  In our particular application, it is computationally challenging if not entirely impractical to solve.  Finding the minimum cost flow in a bipartite graph with $\sim$30 million nodes and $\sim$15 trillions edges is time and resource intensive solution. As such, the practicality of the above algorithm in real world problems is limited. \par
In Algorithm \ref{alg:algorithm1}, cost function represents the distance between panelists from two different panels. Two panelists with very high distance are very unlikely to be assigned to each other even when solved with one single bipartite graph. Assuming this, if we disconnect such distant panelists then the large bipartite graph can be clustered (partitioned) into smaller sub-graphs. A subset of distance features can be used as clustering (partitioning) parameters which is an effective heuristic to localize panelists into smaller clusters. Any residual panelists and their weights can be re-clustered (re-partitioned) by $\mathit{relaxed}$ clustering (partitioning) parameters into sub-graphs to be solved again.\par
These clustering (partitioning) parameters can be relaxed iteratively to give better diversity but with less specificity. It gives any residual panelist a wider pool of panelists to be matched compared to the previous iteration. This approach resembles with $\mathit{search \ query \ relaxation}$ - widely used in information retrieval research area where query parameters are relaxed iteratively to give ample query results in order of specificity.
\par
Intuitively, this $\mathit{iterative \ relaxed \ clustering \ (partitioning)}$ would give similar assignments as Algorithm \ref{alg:algorithm1} with a single large bipartite graph. Clustered (partitioned) smaller sub-graphs can be solved independently on multiple processors, which is ideal for distributed platforms. It makes massive scale panel fusion an embarrassingly parallel problem on any distributed platform with commodity hardware. An obvious additional benefit of splitting the problem into smaller sub-graphs is that the aggregated computing time for the sub-graphs will be significantly less than the time for solving the single dense graph because of the polynomial complexity of the algorithm: $O(n^2 m \log(Cn))$.
\par
$\mathit{Node \ splitting}$ can be also avoided by ignoring any split or fractional panelist assignments and leaving that panelist for the next iteration to be matched with the wider pool of panelists due to relaxed clustering (partitioning).
\par
We observed that features with higher subjective matching priority (e.g, gender versus present of children in household) work very well as clustering (partitioning) parameters and remaining features can be used for distance in the cost function within smaller sub-graphs. Depending on problem definition, geographical location boundaries like zip codes, market areas, states can also be used as a clustering (partitioning) parameters.

\begin{algorithm}
\DontPrintSemicolon
\SetNoFillComment 
\KwData{\\
$U, V \ \mbox{-- disjoint sets of  panelist}$\\
$\qquad w_i,  i \in U $ and $w_j, j \in V$  -- projection weights\\
$\qquad z_i,  i \in U $ and $z_j, j \in V$  -- feature vectors\\

}

\KwResult{\\
$x_{i, j}$ for $i \in U$ and $j \in V$ -- assigned flow for edge $e_{i, j}$
}
\Begin{
Normalize and divide feature vector $z$ into $p$ for clustering (partitioning) and $d$ for distance calculation.\\
\While{ (U and V both have unassigned panelist) }  {
Generate panelistClusters using clustering features $p$, such as, $\{U', \ V'\} \ where \ U' \subset U \ and \ V' \subset V$ \\
\;
\tcc{Parallel for loop}
\ForEach {cluster in panelistClusters}  {
Calculate distance between $\ i \in U' \ and \ j \in V'$ using distance features $d$. \\
Generate bipartite graph $G \ (U', V', E)$ \\ 
\;
Optional step: Prune edges in graph for computational efficiency if needed.
\;
Balance graph by adding dummy node to make sure supplyTotal = demandTotal.
\;
\tcc{cost scaling successive approx.}
$\mathit{ assignedPairs := MCFSolver(bipartite graph) }$ \\
\;
Update U and V with unassigned or partially matched panelists with residual weights. Remove dummy balancing node's assignments. \\
Relax clustering parameter $p$ \\
}
}

\KwRet{$assignedPairs$};
}
\caption{Optimized iterative relaxed fusion algorithm}
\label{alg:optimizedAlgorithm1}
\end{algorithm}

\subsubsection{Pseudocode: }
Algorithm \ref{alg:optimizedAlgorithm1} describes the optimization discussed above. Feature vector $z$ is normalized as before and divided into two feature sets for clustering (partitioning) and distance calculations on Line 2. Lines [3-17] describe iterative relaxation of clustering parameter $p$ which performs a loop until all panelists from either side are completely assigned. Inside this loop, line 4 shows generating clusters (partitions) using features $p$. Lines [6-16] show routine that can run on separate processors in parallel. Line 7 calculates the distance between panelists in same cluster using distance features $d$ and build bipartite graph on line 8. \par
Line 10 explains an optional step to prune edges in the graph if needed. We note that \cite{lee1994very} mentions $2 n \log n$ edges are sufficient for making optimized assignment so pruning the edges can be helpful to speed up computation. Edge pruning is an optional step and is not to be considered the only optimization.   We evaluated pruning edges at random.  In particular, edges below a certain threshold were discarded, disconnecting any node from bipartite graph. Pruning is also helpful to remove unwanted assignments in the solution and can be a reasonable adaptation depending on the domain of a problem.\par
Equation \ref{eq:1} states that total weights in both sets of panelists is exactly the same as they represent the same universe. This does not hold true when we divide panelists into smaller clusters to build sub-graphs. Therefore, we add a dummy \emph{balancing node} on the side of bipartite graph with less supply/demand. The balancing node gets supply/demand equal to the absolute difference between supplyTotal and demandTotal. This process is defined at line 11.\par
Line 12 solves the graph in the same manner as Algorithm \ref{alg:algorithm1}. Line 14 describes the residual process that generates assignment pairs and removes/updates matched panelists and their weights for the next iteration. Balancing nodes,  if added, should be removed from assignment solution set. 
Line 15 relaxes the clustering parameter $p$ so the next iteration generates more diverse but less specific clusters to create a wider panelists pool for assignment.
\bigskip
\bigskip
\bigskip

\subsubsection{Graph balancing: }
There will always be an optimal flow with minimal cost in the graph provided there is enough supply and demand to match and edges with the capacity to transport. We need to make sure that every sub-graph will have an optimal solution by setting the total supply as same as the total demand. If not, we should add a dummy balancing node just to solve the sub-graph optimally. This dummy balancing node $d$ has proxy supply/demand as,

\begin{equation}
    w_d = \sum w_i - \sum w_j
\end{equation}
if $w_d$ < 0, then node d is $\mathit{supply \ node}$; if $w_d$ > 0, then node d is $\mathit{demand \ node}$; 

This dummy balancing node and flows assigned from/to it are removed from the assignment pair solution set. Residual panelists and weights are adjusted after removing the balancing node.

\section{Evaluation}  We evaluated our algorithm on a large scale panel fusion problem from Comscore.  Census-level inferred panel dataset are derived from  measurement data.  The Comscore tagging network is referred as \emph{census panel dataset}. It contains panelist demo identifiers and minutes spent on internet categories across PC, Phone and Tablet platforms. Comscore also maintains \emph{traditional panel dataset}  through recruitment, which contains  similar information. The census panel dataset has the breadth to cover niche audience measurement demands while the traditional panel datasets contain in-depth insights. Combining these two independent panel datasets addresses complex measurement problems.\par
Let $z_i = [z_i^{(1)}, z_i^{(2)}, \dots ]^T$ be the feature vector, with both real valued elements $z_i^{(k)}$ for $k \in R$ and categorical valued elements $z_i^{(k)}$, $k \in C$ otherwise.  Then,
the distance function can be generalized as -
\begin{equation} \label{eqnd}
d(z_i,z_j)=
\sum_{k \in R} \left( z_i^{(k)} -  z_j^{(k)}\right)^2  +
\begin{cases}
\infty & \mbox{if  } z^{(k)}_i \neq z^{(k)}_j \mbox{ for any $k \in C$ } \\
0 & \mbox{otherwise}
\end{cases}.
\end{equation}
In our instantiation, demographic features such as gender, age-group and income are categorical features.  

\emph{Claim 1:}  The relaxed version of the network flow problem solved by Alg. 2 is  optimal if: \emph{i)} the cost per unit flow is defined by Equation \ref{eqnd} and \emph{ii)} node re-balancing is not required. 

We compare our optimized algorithm with the core fusion algorithm containing single graph. We demonstrate results with PC panel datasets only but the same experiments can be performed with Phone and Tablet panel sets as well.  The \emph{Census panel dataset} for PC platform contains approximately 8.7 million panelists while the \emph{traditional panel dataset} has over 450,000 panelists with their demo profiles and online behavior. For simplicity's sake, demo profiles can be categorized as demo categories for \emph{age, gender, household income, ethnicity, race, household size, presence of children in household}. We use the Google OR tool \cite{GoogleOrTool} library for cost scaling successive approximation algorithm to solve assignment problem. This is an open source library with JVM support which makes it easier to use in modern distributed systems like Apache Spark \cite{ApacheSpark}, \cite{Zaharia:2016:ASU:3013530.2934664}.
\par
A single bipartite graph for the above panel datasets would generate around 4 trillion edges, which is computationally expensive to build and solve for the minimum circulation problem. For experiment purpose, we sampled these datasets to 1\% and adjusted projection weights accordingly to represent entire universe. This reduced sample sizes of the \emph{census panel dataset} and the \emph{traditional panel dataset} for PC platform to 87,576 and 4,605 respectively.
\subsubsection{Core fusion algorithm setup}:
We generated a single bipartite graph for these sampled datasets with 403,287,480 edges with the cost between the edges as distance using Equation 4. All features are used for L2 distance calculation. The cost of  edges between panelists with different demo categories, however, was penalized with scalar of 1000 to discourage flow on such edges unless absolutely necessary. This helps to enforce Constraint \Romannum{2} as explained before. We followed Algorithm 1 to generate assignments on a standalone machine.
\subsubsection{Optimized iterative relaxed fusion algorithm setup}:
We followed the optimization described in Algorithm 2. Only real valued features (minutes spent per internet category) are used for L2 distances while demo identifiers are used as clustering (partitioning) parameters. Equation 4 is used for distance calculations. 
In iteration 1, we generated 2858 clusters (partitions) for each unique combination of demo categories and built independent bipartite graphs to solve in parallel. We leveraged the Apache Spark distributed system where each task was given a graph to solve. For simplicity of explanation, we completely relaxed clustering parameter in iteration 2 and combined all demo categories together to create one single graph. Again, the cost of edges between panelists in iteration 2 graph were penalized with scalar of 1000 to discourage flow unless necessary.

\begin{table} [h!]
\scalebox{0.7}{%
\begin{tabular}{ |p{3cm}|p{3cm}|p{5cm}| }
\hline
& Core fusion algorithm & Optimized iterative fusion algorithm \\
\hline
Compute resources & Single node with 150 GB memory, 88 cores  & Spark cluster with 50 executors - 10 GB and 3 cores per executor\\
\hline
End to end execution time & 2.5 Hour & 20 minutes \\
\hline
Cost & x & 2.37x \\
\hline
Total assignments & 92,600 & 92,543 \\
\hline
Assignments within same demo categories & 56,506 (61.02\%) & 58,521 (63.23\%) \\
\hline
Assignments across different demo categories & 36,094 (38.97\%) & 34,022 (36.76\%) \\
\hline
Flow assigned with same demo categories & 56.27\%& 58.60\% \\
\hline
Flow assigned across different demo categories & 43.72\%& 41.39\% \\
\hline
\end{tabular}
}
\caption{Comparative results}
\label{table:1}
\end{table}

Table \ref{table:1} shows comparative results from test runs with the above experimental setup. The core fusion algorithm consumes considerable physical memory on a single node even with just 1\% of sampled data. This proves serious scalability concerns for the core fusion algorithm in practical applications. On the hand, the optimized algorithm is ideal for distributed systems with reasonable compute resources. The optimized version also outperforms the core algorithm for end to end execution time due to the obvious parallelism in the solution. This makes optimized relaxed fusion algorithm easy to adapt on commodity hardware with general purpose distributed systems.\\
The optimal cost of the iterative relaxed fusion algorithm is 2.37 times higher than the core fusion algorithm. We observed that cost with an optimized algorithm decreases as clustering (partitioning) parameters are slowly relaxed with multiple iterations. Comparing the matching results, however, looks very promising.  The core fusion algorithm solution consists of 61.02\% and 38.97\% assignments within same demo categories and across different categories respectively. The optimized algorithm solution has slightly better results with 63.23\% and 36.76\% assignments within same demo categories and across different demo categories. This is due to the stricter partitioning of graph based on demo categories in iteration 1. Overall, an optimized iterative relaxed algorithm comes with an additional cost of a circulation but it is highly practical with general purpose hardware to compute and provides similar assignment results in the end.
\par
Furthermore, we provide results from the full scale PC panel fusion problem described above. Panel fusion of approximately 8.7M panelists from the \emph{census panel dataset} with approximately 450,000 panelists from the \emph{traditional panel dataset} was done with 8 iterations where clustering (partitioning) demo categories are slowly relaxed as shown in Table \ref{table:2}.
Table \ref{table:3} shows number of panelists matched from both \emph{census panel dataset}  and \emph{traditional panel dataset} in every iteration. It demonstrates that panelists from both datasets are completely matched at iteration 7. The number of matched panelists decreases as we relax the clustering (partitioning) parameter. This proves Constraint \Romannum{2} is met, as  more panelists are matched with strict demo profiles.

\begin{table} [h!]
\scalebox{0.66}{%
\begin{tabular}{ |c|l| }
\hline
Iteration & clustering (partitioning) parameters \\
\hline
Iteration 1 & age, gender, ethnicity, household income, race, household size, children \\
\hline
Iteration 2 & age, gender, ethnicity, household income, race, household size \\
\hline
Iteration 3 & age, gender, ethnicity, household income, race \\
\hline
Iteration 4 & age, gender, ethnicity, household income \\ 
\hline
Iteration 5 & age, gender, ethnicity \\
\hline
Iteration 6 & age, gender \\
\hline
Iteration 7 & age \\
\hline
Iteration 8 & No partitioning \\
\hline
\end{tabular}
}
\caption{Clustering (Partitioning) parameters relaxation}
\label{table:2}
\end{table}
\par
\begin{table} [h!]
\scalebox{0.7}{%
\begin{tabular}{ |c|c|c| }
\hline
& Census panel dataset & Traditional panel dataset \\
\hline
Iteration 1 & 7,044,759 & 306,504 \\
\hline
Iteration 2 & 910,327 & 68,191 \\
\hline
Iteration 3 & 128,173 & 14,003 \\
\hline
Iteration 4 & 378,567 & 42,428 \\
\hline
Iteration 5 & 202,914 & 11,241 \\
\hline
Iteration 6 & 56,411 & 9,985 \\
\hline
Iteration 7 & 43,643 & 5,665 \\
\hline
Iteration 8 & 0 & 0 \\
\hline
Total & 8,764,794 & 458,017 \\
\hline
\end{tabular}
}
\caption{Number of matched panelists per iteration}
\label{table:3}
\end{table}
\par
Figure \ref{fig:execution_time} and \ref{fig:clusters} show the decrease in execution time and the number of clusters (partitions) as the iteration increases. Figure \ref{fig:decay} shows decay in unmatched panelists as we advances with iterations. Iteration 7 has 0 unmatched panelists on both sides.
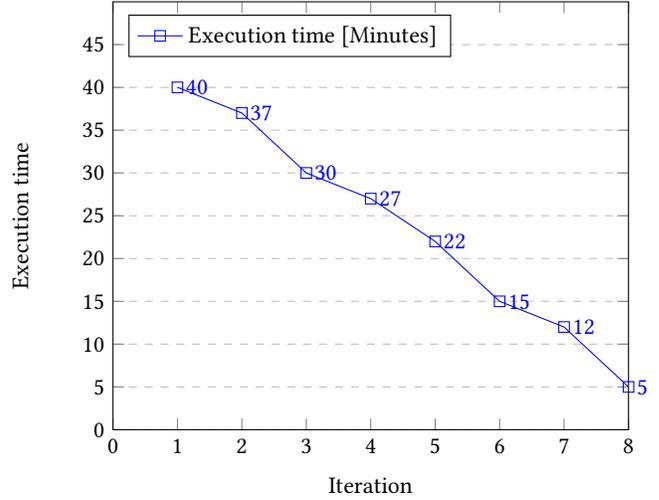
\begin{figure}
\begin{tikzpicture}
\begin{axis} [
	nodes near coords, nodes near coords align={horizontal},
    xlabel={Iteration},
    ylabel={Execution time},
    xmin=0, xmax=8,
    ymin=0, ymax=50,
    xtick={0,1,2,3,4,5,6,7,8},
    ytick={0,5,10,15,20,25,30,35,40,45},
    legend pos=north west,
    ymajorgrids=true,
    grid style=dashed,
]
 
\addplot[
    color=blue,
    mark=square,
    ]
    coordinates {
    (1,40)(2,37)(3,30)(4,27)(5,22)(6,15)(7,12)(8,5)
    };
    \legend{Execution time [Minutes]}
 
\end{axis}
\end{tikzpicture}
\caption{Execution time for every iteration} \label{fig:execution_time}
\end{figure}
\\
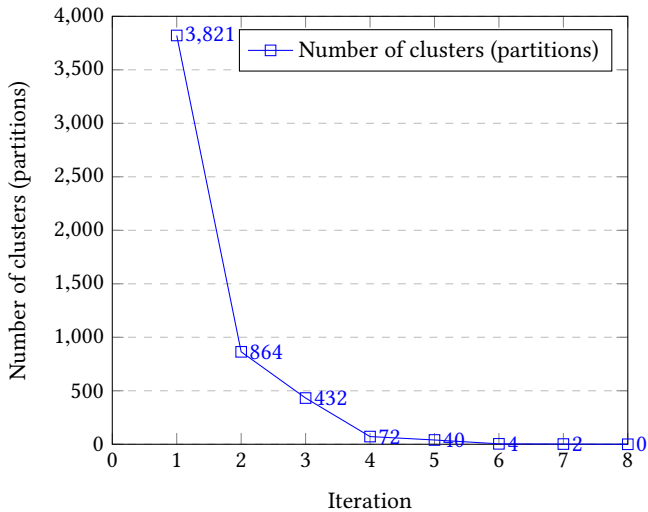
\begin{figure}
\begin{tikzpicture}
\begin{axis} [
	nodes near coords, nodes near coords align={horizontal},
    xlabel={Iteration},
    ylabel={Number of clusters (partitions)},
    xmin=0, xmax=8,
    ymin=0, ymax=4000,
    xtick={0,1,2,3,4,5,6,7,8},
    ytick={0,500, 1000, 1500, 2000, 2500, 3000, 3500, 4000},
    legend pos=north east,
    ymajorgrids=true,
    grid style=dashed,
]
 
\addplot[
    color=blue,
    mark=square,
    ]
    coordinates {
    (1,3821)(2,864)(3,432)(4,72)(5,40)(6,4)(7,2)(8,0)
    };
    \legend{Number of clusters (partitions)}
 
\end{axis}
\end{tikzpicture}
\caption{Number of clusters (partitions) generated in every iteration} \label{fig:clusters}
\end{figure}
\\
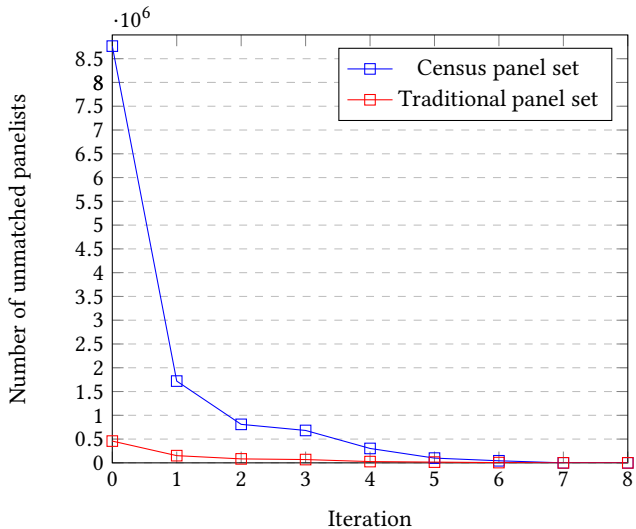
\begin{figure}
\begin{tikzpicture}
\begin{axis} [
    xlabel={Iteration},
    ylabel={Number of unmatched panelists},
    xmin=0, xmax=8,
    ymin=0, ymax=9000000,
    xtick={0,1,2,3,4,5,6,7,8},
    ytick={0,500000, 1000000, 1500000, 2000000, 2500000, 3000000, 3500000, 4000000,  4500000, 5000000, 5500000, 6000000, 6500000, 7000000, 7500000, 8000000, 8500000, 8000000},
    legend pos=north east,
    ymajorgrids=true,
    grid style=dashed,
]
 
\addplot[
    color=blue,
    mark=square,
    ]
    coordinates {
    (0,8764794)(1,1720035)(2,809708)(3,681535)(4,302968)(5,100054)(6,43643)(7,0)(8,0)
    };
    \addlegendentry{Census panel set}
\addplot[
    color=red,
    mark=square,
    ]
    coordinates {
    (0,458017)(1,151513)(2,83322)(3,69319)(4,26891)(5,15650)(6,5665)(7,0)(8,0)
    };
    \addlegendentry{Traditional panel set}
 
\end{axis}
\end{tikzpicture}
\caption{Decay in number of unmatched person in every iteration} \label{fig:decay}
\end{figure}

\subsection{Matching effectiveness}
Validation of panelist matching quality in this problem is very subjective and difficult due to lack of ground truth. We set up an experiment where same panel dataset is used on both sides of bipartite graph for fusion. Any panelist will obviously have the same demographic profile and online behavior when compared to itself. The most ideal match in this case is when same panelist is matched to itself because it will be guaranteed to meet constraints \Romannum{1}, \Romannum{2}. We used the \emph{traditional panel dataset} on both sides of bipartite graph and run our optimized algorithm as explained above. Table \ref{table:4} shows the results for the same. It shows that more than 99\% panelists were matched with themselves which proves that our optimized algorithm generates optimal matches in real world scenarios as well. 

\begin{table} [h!]
\scalebox{0.7}{%
\begin{tabular}{ |c|c|c| }
\hline
& Traditional panel dataset (A) & Traditional panel dataset (B) \\
\hline
Number of panelists & 458,017 & 458,017 \\
\hline
Same panelist matches & 99.37\% (455,127) & 99.37\% (455,127) \\
\hline
Different panelist matches & 0.63\% (2,890) & 0.63\% (2,890) \\
\hline
\end{tabular}
}
\caption{Quality of matches results}
\label{table:4}
\end{table}

This strengthens our belief that using \emph{relaxed distributed minimum cost flow} algorithm guarantees generation of optimal assignments (matches) in real world panel fusion at scale. Comscore runs several independent analysis of using  panel assignments from this algorithm for cross platform campaign evaluations. Results are very much aligned with actual panelists observed across multiple platforms while increasing the scale of measurement for better accuracy. Many niche cross platform campaigns which were difficult to measure with limited observable panelists are now very much possible due to the scale of the of the disparate panel datasets that we can now fuse.
\section{Related Work} \label{sec:relatedwork} While there is limited academic literature on large scale panel fusion problems, network optimization and its applications in real world problems has a very rich history. Network flow problems were first studied by Russian mathematician A.N. Tolstoi in the 1930s to build a railway network in Russia. \cite{ahuja1995applications} discusses various applications of network optimization from the fields of operation research, computer science, medicine, engineering, and applied mathematics. 
\par
Minimum cost circulation is a generalization of a maximum flow problem.  We refer the reader to \cite{goldberg1990finding} for an outline of the problem, and a solution by cost scaling successive approximation. \cite{goldberg1997efficient} and \cite{bunnagel1998efficient} further provide an efficient implementation of same algorithm with heuristics improvements.  This work is closely related to Alg. 1.  This improves practical running time of the algorithm but does not improve worst case complexity of algorithm which is $O(n^2 m \log(Cn))$ where $C$ is a constant that bounds the largest cost in the graph. This algorithm works practically better than other min cost flow algorithms as proved in \cite{goldberg1997efficient}.  

\cite{burkard1999linear} describes different aspects of linear assignment problem and various applications of it. \cite{Soong2001THEAO} discusses panel fusion with unconstrained and constrained statistical matching. The constrained matching is based on a similar transportation problem solved with widely known stepping algorithm. It also demonstrates small scale fusion of TV panelists with magazine/product usage survey. \par

Network flow algorithms are widely used in modern assignment problems across different domains. \cite{butt2013multi} and \cite{zhang2008global} demonstrates effective use of min cost network flow in computer vision problem. \cite{lenz2015followme}, \cite{kitti} extends this work to improve the algorithmic complexity of min cost flow in online vehicle tracking systems. \cite{tian2015text} provides use of min-cost flow network in text detection systems. We want to utilize this cross domain research work on minimum cost flow network to improve computational complexity of offline panel fusion at scale.\par

Our algorithm is aimed at transforming the generic panel fusion problem into a network flow and solving it with massive panel datasets. As discussed in section 3.2, \emph{Iterative relaxation of clustering (partitioning) parameters} provide optimization ideal for distributed systems. Feature selection, normalization, and clustering (partitioning) is very generic so can be tweaked based on problem definition. We evaluated our optimization with sampled datasets and found similar results as solved with a single huge bipartite graph.\par
Our methodology allows panel fusion at scale to be solved efficiently and in an optimal way with commodity hardware.

\section{Summary and Future Work} \label{sec:summary} In this paper, we aim to solve the problem of scalable panel fusion by panelists assignments from two independent datasets. This is an important problem to solve to meet the demands of highly specific but accurate audience measurement. Media is being consumed by a plethora of platforms and devices with various consumption patterns which makes use of a single monolithic panel dataset obsolete. Panel fusion methodology helps to combine such disparate independent panel data sources together to build a comprehensive and cohesive audience panelist dataset. Learning audience behavior from such a fused panel dataset is critical for the success of accurate content and advertisement measurement. We develop a methodology to successfully transform panel fusion and its constraints into a transportation problem and solving it with minimum cost circulation methods. Cost scaling successive approximation algorithm is used to solve minimum cost flow optimally. We provide an optimized iterative relaxation fusion algorithm to solve real world large scale assignment problems. We evaluated our optimization method with Comscore's census and traditional panel datasets. The optimization method provides similar assignments compared to the naive method where all assignments are generated using a single huge graph. Optimized algorithm is computationally very efficient and can easily scale with general purpose distribute platforms. At the same time, our algorithm remains very generic where domain specific clustering (partitioning) methods can be easily applied while core iterative minimum cost circulation methodology remains the same. A generic implementation of this same algorithm running on Apache Spark is successfully used for large scale panel fusion problems in Comscore for various countries, platforms etc. Expanding this same algorithm beyond panel fusion and improving algorithmic complexity to solve generic combinatorial assignments will be our future work.

\bibliographystyle{ACM-Reference-Format}
\balance
\bibliography{paper}

\end{document}